\documentclass[10pt,journal,compsoc, twoside]{IEEEtran}
\pdfoutput=1 

\ifCLASSOPTIONcompsoc
  \usepackage[nocompress]{cite}
\else
  \usepackage{cite}
\fi
\usepackage{amsmath}
\usepackage{footnote}
\usepackage{fancyhdr}
\usepackage{ragged2e}
\usepackage{enumitem}
\usepackage{amsfonts}
\usepackage{booktabs}
\usepackage{multirow}
\usepackage[small]{caption}
\usepackage{bbding}
\usepackage{pifont}
\usepackage{float}
\usepackage{subfigure}
\usepackage{graphicx}
\usepackage{url} 
\begin{document}
%
\title{GCN for HIN via Implicit Utilization of Attention and Meta-paths}
%
%
%
%

\author{Di~Jin, Zhizhi~Yu, Dongxiao~He, Carl~Yang, Philip~S.~Yu and Jiawei~Han
\IEEEcompsocitemizethanks{\IEEEcompsocthanksitem D. Jin, Z. Yu and D. He are with the College of Intelligence and Computing, Tianjin University, Tianjin 300350, China. E-mail: \{jindi, yuzhizhi, hedongxiao\}@tju.edu.cn.
\IEEEcompsocthanksitem C. Yang is with the Department of Computer Science, Emory University, Georgia 30322 USA. E-mail: j.carlyang@emory.edu.
\IEEEcompsocthanksitem P. S. Yu is with the Department of Computer Science, University of Illinois at Chicago, Chicago, IL 60661 USA. E-mail: psyu@uic.edu.
\IEEEcompsocthanksitem J. Han is with the Department of Computer Science, University of Illinois at Urbana-Champaign, Urbana, IL 61801. E-mail: hanj@cs.uiuc.edu.
}

}

\IEEEtitleabstractindextext{%
\begin{abstract}
\justifying
Heterogeneous information network (HIN) embedding, aiming to map the structure and semantic information in a HIN to distributed representations, has drawn considerable research attention. Graph neural networks for HIN embeddings typically adopt a hierarchical attention (including node-level and meta-path-level attentions) to capture the information from meta-path-based neighbors. However, this complicated attention structure often cannot achieve the function of selecting meta-paths due to severe overfitting. Moreover, when propagating information, these methods do not distinguish direct (one-hop) meta-paths from indirect (multi-hop) ones. But from the perspective of network science, direct relationships are often believed to be more essential, which can only be used to model direct information propagation. To address these limitations, we propose a novel neural network method via \emph{implicitly} utilizing attention and meta-paths, which can relieve the severe overfitting brought by the current over-parameterized attention mechanisms on HIN. We first use the multi-layer graph convolutional network (GCN) framework, which performs a discriminative aggregation at each layer, along with stacking the information propagation of direct linked meta-paths layer-by-layer, realizing the function of attentions for selecting meta-paths in an indirect way. We then give an effective relaxation and improvement via introducing a new propagation operation which can be separated from aggregation. That is, we first model the whole propagation process with well-defined probabilistic diffusion dynamics, and then introduce a random graph-based constraint which allows it to reduce noise with the increase of layers. Extensive experiments demonstrate the superiority of the new approach over state-of-the-art methods.
\justifying
\end{abstract}

}

\maketitle

\IEEEdisplaynontitleabstractindextext

%
\IEEEpeerreviewmaketitle


\ifCLASSOPTIONcompsoc
\IEEEraisesectionheading{\section{Introduction}\label{sec:introduction}}
\else
\section{Introduction}
\label{sec:introduction}
\fi

%
%
%
%
\IEEEPARstart{H}{eterogeneous} information networks (HINs)\cite{17}\cite{23}\cite{27}, which involve a diversity of node types and relationships between nodes, can better model and solve many real-world problems than homogeneous networks. For HIN analysis, an important concept is meta-path \cite{25}\cite{26}, which is composed of a sequence of relationships between two nodes. For example, the movie network of IMDB contains three types of nodes, including movies, directors and actors. The relationship between two movies can be described by meta-paths such as Movie-Actor-Movie (MAM) and Movie-Director-Movie (MDM), where MAM denotes the movies starring the same actor, and MDM denotes the movies directed by the same director. 

Network embedding\cite{21}\cite{24}, which aims to learn the distributed representations of nodes in networks, is considered as an effective method for network mining and has been widely studied in homogeneous networks. Recently, researchers have also proposed some methods for HIN embedding, such as random walk-based methods\cite{3}\cite{8} and relation learning based methods\cite{15}\cite{16}, many of which rely on the concept of meta-path. In particular, with the great success of deep learning, graph neural network-based HIN embedding methods (such as HAN \cite{7} and MAGNN\cite{12}) have been proposed very recently. These methods often adopt a hierarchical attention structure, which uses the node-level attention to aggregate information inside each meta-path and utilizes the meta-path-level attention to fuse information of different meta-paths.

While these graph neural network-based methods have achieved great success in HIN embedding, they still suffer from some essential issues. First, while attention has been widely used in fields such as NLP, the use of the complicated hierarchical attention structure may be not so effective in HIN embedding, since there are often little training data available in HINs and information from one network can be hardly transferred to another. In this way, it will be difficult for graph neural networks to train well these hierarchical attentions (particularly for the meta-path-level attention, which is to evaluate the essential importance of different meta-paths), making them hard to really achieve the goal of selecting meta-paths, especially when there is often severe overfitting in practice. At the same time, these existing methods often treat meta-paths with different lengths, such as direct linked meta-paths (e.g., Movie-Director) and indirect linked meta-paths (e.g., Movie-Director-Movie), indistinguishably for information propagation. However, from the perspective of network science, while direct links can propagate information directly, indirect links should propagate information indirectly, and the information propagation on direct links is more essential. Therefore, for meta-paths with lengths longer than one (which makes the paths indirect), it is intuitive that the information should be propagated indirectly rather than directly. Fortunately, we find that graph convolutional network (GCN)\cite{2} itself can partly overcome this limitation. It can realize that direct linked meta-paths propagate information directly at each layer, and indirect linked meta-paths propagate information indirectly via the stacked layers of deep neural networks. More importantly, it has already encoded the information of all meta-paths via the multi-layer propagation in an implicit way. However, GCN does not distinguish the importance of information from different meta-paths in both its propagation and aggregation processes, which makes it not directly suitable for HIN embedding.

To utilize the advantages of GCN of implicitly encoding all meta-paths as well as overcome the difficulty of distinguishing their importance in an effective way, we propose a novel \textbf{G}CN-based approach for heterogeneous information network via \textbf{I}mplicit utilization of \textbf{A}ttention and \textbf{M}eta-paths, referred to as GIAM. We first introduce a naive model. It uses the direct linked meta-paths alone for information propagation, and utilizes a new aggregation mechanism for each-layer, along with the stacked-layer propagation, to implicitly achieve the role of attention for selecting meta-paths. In this way, we realize the selection of different meta-paths in GCN itself (rather than using attention directly which may lead to overfitting). Meanwhile, we make an effective refinement. That is, we replace the spectral filter of GCN from the symmetric normalized graph Laplacian to an equivalent asymmetric one, along with removing activation, modeling the propagation with continuous Markov dynamics. We then introduce an effective \textbf{R}andom graph-based \textbf{P}ropagation \textbf{C}onstraint  principle, namely RPC, i.e., if a propagation path on the given network is no better than that on the corresponding random graph, there is no reason to continue this path propagation, which makes the whole propagation process more effective via filtering more impurity information. 

To summarize, the main contributions of this paper are as follows:
\begin{itemize}[leftmargin=*]
\item We find that, the hierarchical attention structure adopted by many HIN-specific graph neural networks is hard to really achieve the function of essential selections of meta-paths (due to severe overfitting); and meanwhile, they do not distinguish one-hop and multi-hop meta-paths in the propagation process.
\item We propose a new approach to solve these problems. It uses only direct linked meta-paths for direct propagation and realizes indirect propagation by stacking layers of direct propagations. We distinguish the importance of information from different meta-paths (in this process) via effective algorithmic mechanisms rather than using attentions directly.
\item Extensive experiments on different network analysis tasks demonstrate the superiority of the proposed new approach over some state-of-the-arts.
\end{itemize}

The rest of the paper is organized as follows. Section 2 introduces a motivating example. Section 3 gives the problem definitions and introduces GCN. Section 4 proposes the new approach for HIN embedding. In Section 5, we conduct extensive experiments. Finally, we discuss related work in Section 6 and conclude in Section 7.


\section{A Motivating Example}
To verify whether using meta-path-level attention can effectively evaluate the importance of different
\renewcommand\arraystretch{1.6}
\begin{table}[!ht]
	\centering
	\caption{\label{table:attention}The performance of HAN and MAGNN of using (and not using) meta-path-level attention, as well as our new approach GIAM on IMDB and DBLP. 'Y'  denotes the method of using meta-path-level attention and 'N'  not. '-'  denotes our new idea of using algorithmic mechanisms rather than attention to learn relationships of meta-paths. Attention distribution is denoted by the learned weights of importance of different meta-paths.}
	\begin{small}
	\resizebox{\linewidth}{!}{
		\begin{tabular}{|c|c|c|c|c|c|c|}
			\hline
				{{Datasets}} &{{Meta-paths}} & {{Models}} & {{Attention}} & {{Attention distribution}} & {{Macro-F1}} & {{Micro-F1}} \\ 
			\hline
				\multirow{5}*{IMDB} & \multirow{5}*{\shortstack{MDM\\MAM}} & \multirow{2}*{HAN} & {Y} &
				{[0.78, 0.22]} & {57.67} & {57.79}
				 \\ \cline{4-7}
				{} & {} & {} & 	{N} &
				{[0.50, 0.50]} & {58.93} & {59.02} 
				 \\ \cline{3-7}
				{} & {} & \multirow{2}*{MAGNN} & {Y} & {[0.57, 0.43]} & {57.60} & {57.72}
				 \\ \cline{4-7}
				{} & {} & {} & {N} &
				{[0.50, 0.50]} & {58.30} & {58.50}
				 \\ \cline{3-7}
				{} & {} & {GIAM} & {-} & {-} & {59.58} & {59.86}
				\\ \cline{1-7}
				\multirow{5}*{DBLP} & \multirow{5}*{\shortstack{APA\\APVPA\\APTPA}} & \multirow{2}*{HAN} & {Y} &
				{[0.258, 0.736, 0.006]}
				& {92.69} & {93.20}
				 \\ \cline{4-7}
				{} & {} & {} & 	
				{N} & {[0.333, 0.333, 0.333]}
				& {92.47} & {93.04}
				 \\ \cline{3-7}
			{} & {} & \multirow{2}*{MAGNN}  & {Y} & {[0.022, 0.969, 0.009]}
				& {93.19} & {93.67}
				\\ \cline{4-7}
			{} & {} & {}  & {N} & {[0.333, 0.333, 0.333]}
				& {90.42} & {91.08}
				\\ \cline{3-7}
            {} & {} & {GIAM} & {-} & {-} & {93.63} & {94.10}
				 \\ \hline
	    \end{tabular}
	}
	\end{small}
\end{table}
meta-paths, we conduct experiments on two widely-used heterogeneous information networks, i.e., IMDB and DBLP. We select three graph neural network-based HIN embedding methods, i.e., HAN, MAGNN and our new approach GIAM (which will be introduced in Section 4 below). Since HAN and MAGNN require a candidate meta-path set, and our GIAM can also support this option, we use the same choices according to the existing work \cite{7}\cite{12}, i.e., \{MDM, MAM\} for IMDB ('M/D' stands for Movie/Director and 'A' stands for Actor) and \{APA, APVPA, APTPA\} for DBLP ('A/P' stands for Author/Paper and 'V/T' stands for Venue/Term), which are often believed to be the essential meta-paths for node classification in networks. We compare HAN (and MAGNN) of using and not using meta-path-level attention, as well as our new idea (GIAM) of using algorithmic mechanisms (rather than attention) to learn relationships of meta-paths. We first get each method's embedding on each dataset (according to the experimental settings in Section 5), and then feed them to SVM classifier with different radios (i.e., 5\%-80\%) of supervised information. We report the average accuracy 
over these radios, in terms of Macro-F1 and Micro-F1, as shown in Table 1; and show the detailed accuracy on each radio of the supervised information in Appendix.

As shown, on IMDB, it is surprising that, the methods (HAN and MAGNN) of using meta-path-level attention are always no better than those of not using it. Concretely, for HAN of using meta-path-level attention, it is easy to obtain the staple attention distribution, where one dominant meta-path has the dominated attention value (i.e., the distribution [0.78, 0.22] on \{MDM, MAM\}). Though this seems to achieve a well evaluation of the importance of different meta-paths, the accuracy is surprisingly reduced. This may be mainly due to overfitting, preventing the method from really selecting correct meta-paths. Differently, MAGNN with meta-path-level attention is easy to get the smooth attention distribution, i.e., [0.57, 0.43] on \{MDM, MAM\}. While the learned attention values differ slightly, the accuracy is still not improved when comparing with that of not using attention. On the other hand, on DBLP, the methods (HAN and MAGNN) of using meta-path-level attention perform slightly better than those of not using it. Since these models on DBLP can be trained much better with a high accuracy (compared with those on IMDB), they may relieve overfitting and make attention effective to some extent. But anyway, in both these two settings, our new approach GIAM of using the specially designed algorithmic mechanisms (rather than attention) to learn relationships of meta-paths stably performs the best.

\begin{figure}[htbp]
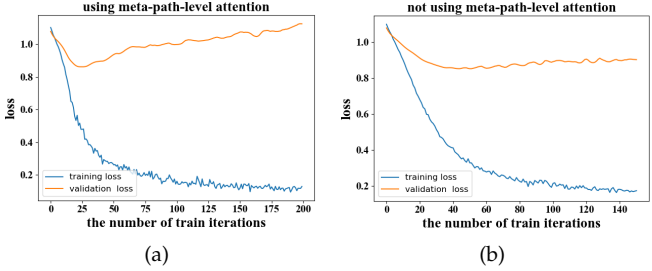

\vspace{-0.3cm}
\setlength{\abovecaptionskip}{-0.05cm} 
\centering
\setlength{\belowcaptionskip}{-0.5cm} 
\subfigure[]{
\label{with atten}
\includegraphics[width=0.47\linewidth]{with_attention.pdf}
}
\subfigure[]{
\label{without atten}
\includegraphics[width=0.47\linewidth]{without_attention.pdf}
}   

\caption{\label{Attention}The results of training loss (and validation loss) as a function of the number of train iterations by using HAN on IMDB. (a) shows the result of using meta-path-level attention and (b) shows that of not using meta-path-level attention.}
\end{figure}

To further verify whether overfitting is the main reason that meta-path-level attention does not help evaluate the importance of different meta-paths effectively, we conduct extra experiments on IMDB by using HAN as an example. We show the training loss (and validation loss) as a function of the number of train iterations. Fig. \ref{with atten} shows the result of HAN of using meta-path-level attention, and Fig. \ref{without atten} shows that of not using meta-path-level attention. As shown, when using meta-path-level attention, with the decrease of the training loss, the validation loss first decreases but then increases significantly, which is a highly overfitting phenomenon. Differently, the overfitting issue is relative slight when not using the meta-path-level attention. This partly validates that the meta-path-level attention may not be able to achieve well the essential selection and evaluate the importance of different meta-paths, especially when the model is hard to be trained well (which is often the real life in many network analysis tasks).

\section{Preliminaries}
We first introduce the problem definition, and then discuss GCN which serves as the base of our new approach.
\subsection{Problem Definition}

\textbf{Definition 1. Heterogeneous Information Network}. A heterogeneous information network is defined as a network $G(V, E, F, R, \phi, \varphi)$, where $V$ represents the set of multiple types of nodes, $E$ the set of multiple types of edges, and $F$ and $R$ the set of node and edge types. Each node ${u}\in{V}$ is associated with a node type mapping function $\phi: {V} \rightarrow {F}$, and each edge ${e}\in{E}$ is associated with an edge type mapping function $\varphi: {E} \rightarrow {R}$. $G$ is defined as a heterogeneous information network when $|F| + |R| > 2$.

\textbf{Definition 2. Adjacency Matrix of Heterogeneous Information Network}. Inspired by homogeneous network, we define the adjacency matrix of heterogeneous information network $G$ as $A=(a_{uv})_{n \times n}$, where $a_{uv}= 1$ if there is an edge between nodes $u$ and $v$, or 0 otherwise, and $n = |V|$ the number of nodes. Thus, the degree distribution of $G$ can be defined as $D$ = diag$(d_1,...,d_n)$, where $d_u =  \sum\nolimits_{v} a_{uv}$, i.e., we sum up the number of edges associated with node $u$.

\textbf{Definition 3. Meta-path}. A meta-path $m$ is defined as a path in the form of ${F_1}\stackrel{R_1}{\longrightarrow} {F_2}\stackrel{R_2}{\longrightarrow}...\stackrel{R_l}{\longrightarrow}{F_{l+1}}$(abbreviated as $F_{1}F_{2}\cdots{F_{l+1}}$), where $F$ and $R$ are node and edge types, respectively. It represents a compositional relation between two given node types.

\textbf{Definition 4. Meta-path-based Neighbors}. Given a meta-path $m$ of a heterogeneous information network, the meta-path-based neighbors $N_u^m$ of node $u$ are defined as the set of nodes which connect with node $u$ via meta-path $m$. Note that $N_u^m$ include $u$ itself if $m$ is symmetric.

\textbf{Definition 5. Heterogeneous Information Network Embedding}. Given a heterogeneous information network $G$, this task is to learn the $d$-dimensional distributed representation $H\in\mathbb{R}^{|V| \times d}(d \ll |V|)$ that is able to capture rich structural and semantic information involved in $G$. 

\subsection{Graph Convolutional Network}
Spectral graph convolutional neural networks (GCN) is proposed by Bruna \emph{et al.}\cite{14} to analyze the graph data. It defines spectral graph convolution as the product of a signal $x$ and a filter $g_ {\theta}$ = diag$(\theta)$, where $\theta$ is a vector in the Fourier domain. Following this, the spectral graph convolution can be performed as $g_{\theta} \star x = U{g_\theta} {U^T}x$, where ${U^T}x$ is the graph Fourier transform of $x$ and $U$ the matrix of eigenvectors of the normalized graph Laplacian $L$ defined as $L =
I-D^{-{1/2}}AD^{-{1/2}}=U {\Lambda} U^T$ (where $I$ is the identity matrix and ${\Lambda}$ the diagonal matrix of eigenvalues). Since the calculation of eigenvalue decomposition of $L$ in a large graph is very expensive, Defferrard \emph{et al}.\cite{1} suggest to use the $k$-th order Chebyshev polynomial expansion to approximate $g_\theta$, represented as $g_{\theta^{\prime}} \approx  \sum\nolimits_{k=0}^K \theta_k^{\prime}T_k( \widetilde{\Lambda})$, where $\theta_k^{'}$ is the $k$-th Chebyshev coefficient and $ \widetilde{\Lambda} = {(2 /\ { \lambda_{max}}}) \Lambda - I $ (${\lambda}_{max}$ is the largest eigenvalue of $L$). By substituting it into $g_ \theta \star x = U{g_ \theta}{U^T}x$, and adopting ${\widetilde{L} = {(2 /\ {\lambda_{max}}} ) L - I }$, we have ${g_{\theta^{\prime}} \star x \approx  \sum\nolimits_{k=0}^K \theta_k^{\prime}T_k( \widetilde{L})x}$.

Furthermore, Kipf \emph{et al}. \cite{2} propose to use $k = 1$ and ${\lambda_{max}}=2$ to get a simplified graph convolution operation of GCN, represented as $g_{\theta} \star x \approx \theta(I+D^{-{1/2}}AD^{-{1/2}})x$. In addition, by introducing an effective renormalization $\hat{A}= \widetilde D^{-{1/2}}\widetilde A \widetilde D^{-{1/2}}$ (where ${\widetilde A = A+I}$ and $\widetilde D$ = diag$(\widetilde d_1,...,\widetilde d_n)$ with $\widetilde d_u = \sum_v{\widetilde a_{uv} }$), the classic two-layer GCN can then be defined as:
\begin{equation}
\widehat Y=\operatorname{softmax}(\widehat A \operatorname{ReLU}(\widehat A H^{(0)}W^{(0)})W^{(1)})
\end{equation}
where $H^{(0)}$ is the node feature matrix, $W^{(0)}$ (and $W^{(1)}$) the weight parameter of neural networks, 
and $\widehat Y$ the final output for the assignment of node 
\begin{figure}[htbp]
\setlength{\abovecaptionskip}{-0.03cm}
	\centering
 	\vspace{-0.3cm}
    \includegraphics[width=0.8\linewidth]{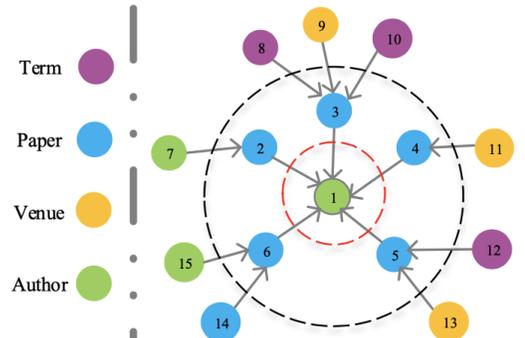}\\
	\caption{\label{GCN_graph}An illustrative example of using GCN on a heterogeneous information network DBLP. The inner (red) circle represents the first layer and the outer (black) circle the second layer.}
\end{figure}
\begin{figure*}[t]
\setlength{\abovecaptionskip}{-0.05cm}
\setlength{\belowcaptionskip}{-0.3cm}
	\centering
 	\vspace{-0.3cm}
    \includegraphics[width=0.98\linewidth]{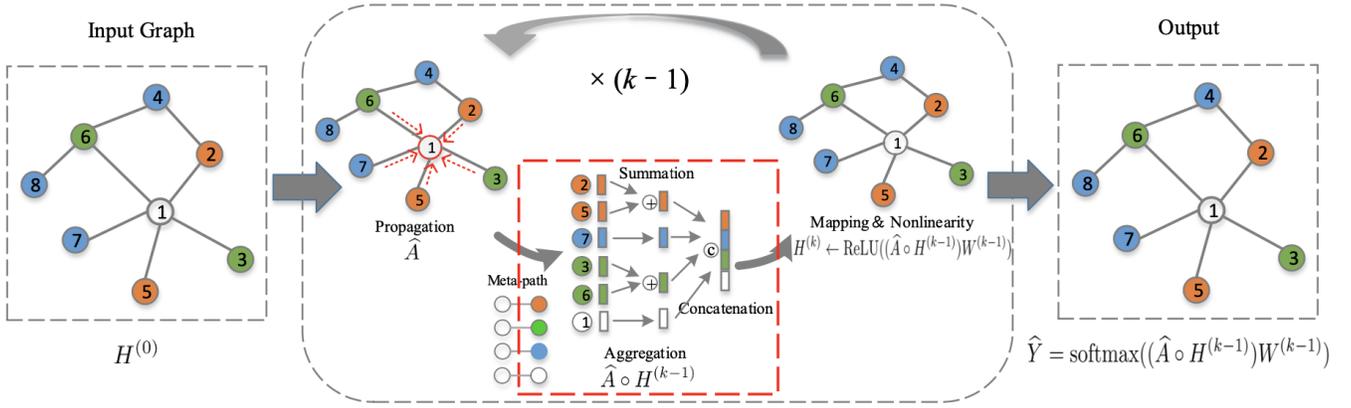}\\
	\caption{\label{Framework1} The structure of the naive model. It propagates and aggregates the information of direct linked meta-path-based neighbors repeatedly via $k$ layers. The part in red box is the \emph{core} content.}
\end{figure*}
labels. While GCN works very well on homogeneous networks, it is not directly suitable for heterogeneous information networks with different types of nodes and edges \cite{32}.

We now analyze the advantages and disadvantages
of using GCN on heterogeneous information networks (by taking DBLP with four types of nodes: author, paper, venue and term as an example). As shown in Fig. \ref{GCN_graph}, in the first layer of GCN (the inner circle in the figure), we can realize the direct information propagation via direct linked meta-paths (e.g., Paper-Author). By stacking the second layer (the outer circle), we can achieve the indirect information propagation of meta-paths with length 2, such as meta-paths Term-Paper-Author and Venue-Paper-Author, with the help of stacked direct linked meta-path propagation. By adopting a multi-layer GCN, we can then realize that the direct linked meta-paths propagate information directly while indirect link meta-paths propagate information indirectly, along with covering meta-paths with different lengths. However, for heterogeneous information networks, GCN often treats the information from different meta-paths equally in the process of both propagation and aggregation, without distinguishing the difference of their importance, which is a challenge and correctly the main limitation we will overcome in this work.


\section{Methodology}
We first propose a naive model to solve the issue of GCN on heterogeneous information networks (HINs), then refine the model by introducing a continuous Markov propagation process, and finally give some optional tricks in implementation.

\subsection{The Naive Model}
In the first model, we use the classic multi-layers GCN as a basic framework, and then introduce a discriminative mechanism to aggregate information from the neighbors with direct linked meta-paths. The structure of this model is illustrated in Fig. \ref{Framework1}. 

The novel aggregation mechanism consists of two parts, including the aggregation of instances under the same meta-path (which we call the intra aggregation) and the aggregation of different meta-paths (which we call the inter aggregation). Specifically, in the intra aggregation, we adopt the same summation as GCN to aggregate the information from the same direct linked meta-path-based neighbors. Mathematically, let $\tau: (u,v) \rightarrow m \in M$ be the meta-path mapping function, where $M$ is the set of direct linked meta-paths. It inputs a node pair $(u,v)$, and outputs a variable $m$ which indicates the direct linked meta-path between nodes $u$ and $v$. Simultaneously, let $h_u^{(k-1)}$ be the embedding of node $u$ at the ($k$-1)-th layer, and $h_u^{(0)}$ the node’s feature vector. Then, for each $u \in V$, its embedding of the direct linked meta-path $m$ at the $k$-th layer $e_u^{(m,k)}$ can be updated as:
\begin{equation}
e_ {u} ^ {(m, k)} =  \sum\limits_{v \in N_u }   \delta ( \tau (u,v),m)(\widetilde d_u \widetilde d_v)^{1\over 2 }h_{v}^{(k-1)} , {\forall}  m \in M 
\end{equation}
where $\widetilde d_u$ is the degree of node $u$ of $G$ with self-edges (as defined in (1)), $N_u$ is the set of direct linked meta-path-based neighbors of node $u$, and $\delta(\cdot, \cdot)$ a Kronecker delta function that only allows nodes with the direct linked meta-path $m$ to node $u$ to be included. Since there are $|M|$ different direct linked meta-paths, then for each node $u$, we will get $|M|$ meta-path-type embeddings. In this case, we adopt another aggregation function, i.e., concatenation $\mathop{\parallel}$, to aggregate the embeddings of different direct linked meta-paths, 
that is:
\begin{equation}
g_{u}^{(k)} =  \mathop{\parallel}_{m \in M}e_{u}^{(m,k)}  
\end{equation}
With the obtained $g^{(k)}_u$, the $k$-th layer embedding of node $u$ can then be given by using a mapping function along with a non-linear transform as: 
\begin{equation}
h_{u}^{(k)} =  \sigma (g_u^{(k)}  \cdot W^{(k-1)} ) 
\end{equation}
where $W^{(k-1)}$ is the mapping matrix and $\sigma(\cdot)$ the non-linear activation function. To simplify expression, we use a new operator '$\circ$' to denote the incorporation of the above two types of aggregations on matrices. Then, the matrix form of the $k$-th layer embeddings can be defined as:
\begin{equation}
 H^{(k)} =  \sigma((\widehat A \circ H^{(k-1)}  ) W^{(k-1)}) 
\end{equation}

\begin{figure}[htbp]
\setlength{\belowcaptionskip}{0.1cm} 
	\centering
    \includegraphics[width=0.8\linewidth]{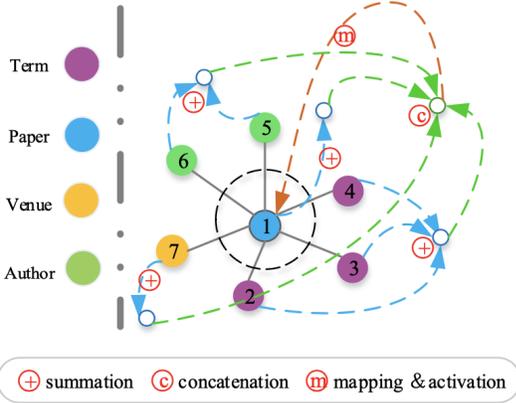}\\
	\caption{\label{naive_model}An illustrative example of using the naive model on DBLP. The blue arc (of using summation) represents the aggregation of information from the same type of neighbors linked by a one-hop meta-path, the green arc (of using concatenation) denotes the aggregation of information from different one-hop meta-paths, and the brown arc (of using the neural network mapping and activation) denotes the selection of different meta-paths by utilizing the inherent algorithm mechanism (i.e., implicit utilization of attention).}
\end{figure}

To better understand how this naive model distinguishes the importance of information from different meta-paths during both propagation and aggregation, we give a brief explanation on a heterogeneous information network (DBLP) as an example. As shown in Fig. \ref{naive_model}, in each layer, we use the direct linked meta-paths within the black circle to propagate information. We adopt the summation to aggregate information from each type of neighbors linked by the same one-hop meta-path (e.g., Author-Paper), and use concatenation to aggregate information from different one-hop meta-paths (e.g., Author-Paper and Term-Paper), and then feed it to the neural network. This is to distinguish the importance of information from different meta-paths in an \emph{implicit} and \emph{indirect} way, i.e., utilizes the new discriminative aggregation as well as the mapping function of neural networks, rather than using attention directly. Furthermore, we extend the propagation range by stacking layer by layer, and then realize the distinction of meta-paths with different lengths (e.g., Author-Paper-Term and Author-Paper-Term-Paper), with the help of the interaction of the multi-layer propagation of the one-hop meta-paths as well as the bi-level aggregation mechanism in each-layer.
\begin{figure*}[t]
\vspace{-0.3cm}
\setlength{\abovecaptionskip}{0.1cm}
\setlength{\belowcaptionskip}{-0.3cm}
	\centering
    \includegraphics[width=0.98\linewidth]{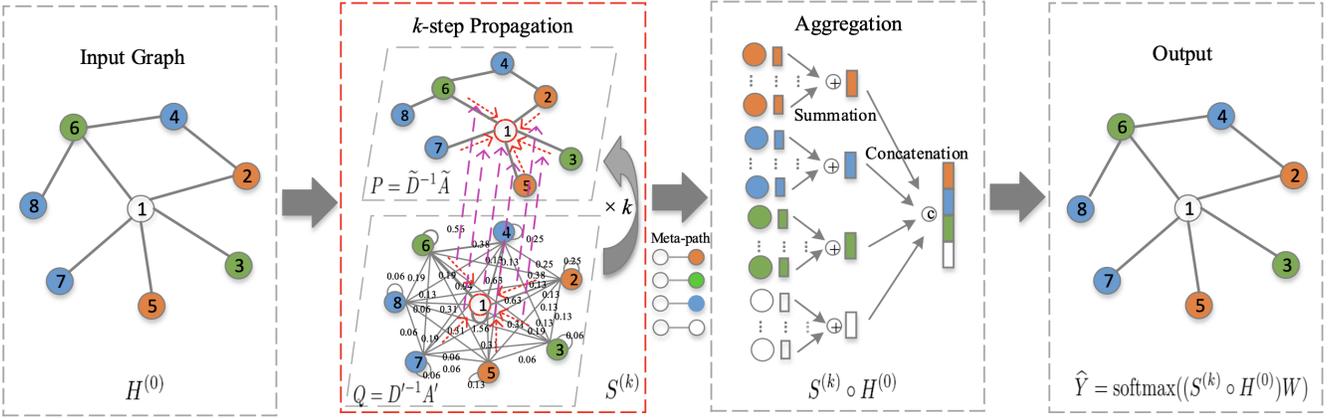}\\
	\caption{\label{Framework2}The structure of the model with constrained Markov propagation. The part in the red box is the \emph{core} improvement and relaxation compared to the naive model.}
\end{figure*}

In fact, while this naive model seems to be able to cover different meta-paths as well as distinguish their importance in both propagation and aggregation in an ideal way, it, however, possesses an inherent limitation, i.e., many nodes do not have the same (or complete) types of one-hop meta-paths due to the sparsity of HIN, making an effective concatenation in this new aggregation process difficult. Take DBLP as an example, some paper nodes may not have links under meta-path Paper-Author while some other nodes may not have links under Paper-Term. In this case, we cannot achieve the alignment of these nodes’ embeddings after concatenation. So, one can only use non-informative vectors (e.g., vectors with all 1 or 0) to fill in these missing types to make them complete. This, however, significantly lowers the performance of the model especially when stacking multi-layers.

\subsection{The Improved Model}

To overcome the limitation of the naive model, we introduce an effective relaxation and improvement. That is, we first perform a $k$-step propagation, and then the discriminative aggregation. In the new propagation process, we replace the spectral filter of GCN from the symmetric graph Laplacian to an equivalent asymmetric one, and then remove activation, in order to make it a continuous Markov dynamics. We then introduce a random graph-based cut mechanism to constrain its free expansion, enabling the propagation to escape from including too many harmful information with the increase of layers. The structure of this model is illustrated in Fig. \ref{Framework2}. In the following, we will introduce 
it from two perspectives, i.e., probabilistic propagation and  discriminative Aggregation.

\subsubsection{Probabilistic Propagation}
First we refine the propagation process of GCN. We adopt an asymmetric normalized graph Laplacian $P = \widetilde D^{-1}\widetilde A$, which is also called the Markov transition probability matrix, as the filter to perform propagation, where
$\widetilde A = A + I$ ($A$ is the adjacency matrix of $G$ and $I$ the identity matrix), and $\widetilde D$ = diag$(\widetilde d_1,...,\widetilde d_n)$ with $\widetilde d_u =  \sum\nolimits_{v} \widetilde a_{uv}$. According to spectral graph theories \cite{11}, $P$ has the same spectrum range with the original spectral filter $\widehat A$ of GCN (defined in (1)), and thus possesses the same ability of serving as a low-pass-type filter for propagation. Meanwhile, we remove activation functions on all layers expect for the output layer (that uses softmax), which will not decrease the model’s performance, as guaranteed by \cite{11}. Then, these two steps make the propagation a continuous Markov dynamics process. The new propagation rule can be defined as: 
\begin{equation}
P^{(k)} =  P^{(k-1)}\cdot P
\end{equation}
where $P^{(0)}=I$.

On the other hand, the above propagation process in graph convolution can be also taken as a $k$-step Markov random walk from the perspective of probabilistic diffusion. Formally, given a heterogeneous information network $G$, the transition probability from nodes $u$ to $v$ within one step random walk can be formulated as:
\begin{equation}
p_{uv}={\widetilde a_{uv}\over \sum_{r}{\widetilde a_{ur}} }
\end{equation}
Then, after walking $k$ steps, the transition probability from nodes $u$ to $v$ can be calculated iteratively by:
\begin{equation}
z_{uv}^{(k)}= \sum_{r=1}^n z_{ur}^{(k-1)} p_{rv }
\end{equation}
where $z_{uu}^{(0)}=1$ and $z_{uv}^{(0)}=0$, for $u \neq v$. The above process can also be taken as a matrix form as:
\begin{equation}
Z^{(k)} = Z^{(k-1)} \cdot P \quad s.t.,\quad Z^{(0)}=I
\end{equation}
where the $k$-step transition probability matrix $Z^{(k)}$ equals to the propagation matrix $P^{(k)}$ in (6) in graph convolution. More interestingly, according to spectral graph theories \cite{30}, the number of steps of random walk in the range of entering and exiting times of the $c$-th local mixing state (of this Markov dynamics) can show the clearest $c$ categories structure. So, this new probabilistic perspective brings a byproduct that we can evaluate the optimal number of propagation layers of graph convolution. To be specific, given a network $G$ with the Markov matrix $P$, the local mixing times of random walks on it can be estimated by using the spectrum of its corresponding Markov generator $M = I-P$, where $M$ is positive semi-definite and has $n$ non-negative real-valued eigenvalues ($0= \lambda _1 \leq\lambda _2 \cdots \leq\lambda _n\leq2$). Let $T_c^{ent}$ and $T_c^{ext}$ be the entering and exiting times of the $c$-th local mixing state, we have $T_c^{ext}= {1\over {\lambda _c}}(1+o(1))$. Reasonably, we can use the exiting time of the ($c$+1)-th local mixing state to estimate the entering time of the $c$-th local mixing state, which can be represented as $T_c^{ent}=T_{c+1}^{ext}= {1/{\lambda _{c+1}}}$. Then, the calculated $T_c^{ent}$ and $T_c^{ext}$ can be taken as the floor and ceiling of the optimal number of propagation layers for a $c$-classification problem.

However, first, it is too time consuming to calculate the eigenvalues for determining the number of propagation layers, which often needs $O(n^3)$ time. Second, even in the expected range of the optimal number of layers, the propagation will still introduce impurity information inevitably, which will also decrease the convolution’s performance. To further overcome those drawbacks, we introduce \emph{the new \textbf{RPC} principle}, i.e., if a propagation path on a given network (with clusters) is no better than that on its corresponding random graph, we will have no reason to continue this propagation path. This will not only enable the propagation to filter more noise information, but also make it not so sensitive to the number of layers (which may be set a relative large value, e.g., 10). To be specific, given a heterogeneous information network $G = (V, E)$, we first calculate its corresponding random graph $G' = (V, E')$ which has the same node degree distribution with $G$ while contains none structural information for classification. We adopt the popular null model of modularity \cite{31} that describes random graphs by rewiring edges randomly among nodes with given node degrees, which is correctly suitable for this work. Let $\widetilde A = (\widetilde a_{uv})_{n\times{n}}$ be the adjacency matrix of $G$ with self-edges, and $\widetilde D$ = diag$(\widetilde d_1,...,\widetilde d_n)$ the degree matrix with $\widetilde d_u =  \sum\nolimits_{v} \widetilde a_{uv}$. Then, based on this null model, the expected number of links (or expected link weight) between nodes $u$ and $v$ can be written as: 
\begin{equation}
a'_{uv}={\widetilde d_u \widetilde d_v\over\sum\nolimits_{r=1}^n {\widetilde d_{r} }}
\end{equation}
which forms the adjacency matrix $A' = (a'_{uv})_{n\times{n}}$ of $G'$. On this random graph, the one step transition probability from nodes $u$ to $v$ can be written as:
\begin{equation}
q_{uv}={a'_{uv}\over \sum_{r}{a'_{ur}}}
\end{equation}
Using it as a constraint on each step of the random walk on $G$, we then get a constraint Markov dynamics. That is, the transition probability from nodes $u$ to $v$ after $k$ steps of the constraint walk, i.e., $s_{uv}^{(k)}$, can be calculated iteratively by:
\begin{equation}
\begin{split}
&s'^{(k)}_{uv}=\operatorname{max}( \sum_{r=1}^n s^{(k-1)} _{ur}p_{rv}-\sum_{r=1}^ns _{ur}^{(k-1)}q_{rv},0 )\\
&\qquad\qquad\qquad s _{uv}^{(k)}= {{ s'^{(k)} _{uv}}\over{ \sum\nolimits_{r=1}^n s'^{(k)} _{ur}}}\\
\end{split}
\end{equation}
where $\sum\nolimits_{r=1}^n s _{ur}^{(k-1)}p_{rv}$ denotes the $k$-step transition probability from nodes $u$ to $v$ on $G$ while $\sum\nolimits_{r=1}^n s _{ur}^{(k-1)}q_{rv}$ the probability on the corresponding random graph $G'$, after $k$-1 steps of the constraint walk. We remove negative values of $s_{uv}^{(k)}$ and normalize it after each step (since the probability distribution should be non-negative and sum to 1). Then, let $S^{(k)} = (s_{uv}^{(k)})_{n \times n}$, $P = (p_{uv})_{n \times n}$, $Q = (q_{uv})_{n \times n}$ and $D_s$ = diag$(d_{s1},...,d_{sn})$ with $ d_{su} =  \sum\nolimits_{v}s_{uv}$, the above process can be rewritten in the matrix form as:
\begin{equation}
\begin{split}
&S'^{(k)} = \operatorname{max}(S^{(k-1)} \cdot P- S^{(k-1)} \cdot Q, 0) \\
&\qquad\qquad S^{(k)} = D_s^{-1} \cdot S'^{(k)}\\
\end{split}
\end{equation}
Finally, we derive the $k$-step transition probability matrix $S^{(k)}$ based on the constraint Markov dynamics, which is to serve as a better propagation matrix for graph convolution. 

To illustrate how the propagation matrix based on the unconstrained (and constrained) Markov dynamics changes 
\begin{figure}[htp]
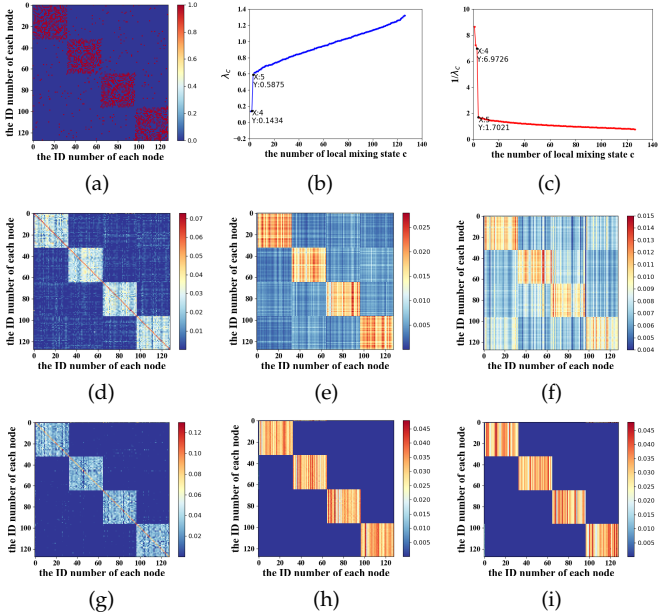

\centering
\subfigbottomskip=0.3pt
\setlength{\abovecaptionskip}{0.13cm}
\subfigure[]{
\label{example_1}
\includegraphics[width=0.293\linewidth]{example_1.pdf}
}
\subfigure[]{
\label{example_2}
\includegraphics[width=0.312\linewidth]{example_2.pdf}
}
\subfigure[]{
\label{example_2}
\includegraphics[width=0.312\linewidth]{example_3.pdf}
}

\subfigure[]{
\includegraphics[width=0.305\linewidth]{unconstraint_2.pdf}
}
\subfigure[]{
\includegraphics[width=0.305\linewidth]{unconstraint_6.pdf}
}
\subfigure[]{
\includegraphics[width=0.305\linewidth]{unconstraint_10.pdf}
}

\subfigure[]{
\includegraphics[width=0.305\linewidth]{constraint_2.pdf}
}
\subfigure[]{
\includegraphics[width=0.305\linewidth]{constraint_6.pdf}
}
\subfigure[]{
\includegraphics[width=0.305\linewidth]{constraint_10.pdf}
}
\caption{An example illustrating that the propagation matrix changes with increasing the number of propagation layers based on the unconstrained (and constrained) Markov dynamics. (a) shows a simple Newman artificial network, (b) the spectrum of its Markov generator, and (c) the exiting (and entering) time of each local mixing state. (d), (e) and (f) show the propagation matrices after 2, 6 and 10 layers of the unconstrained Markov propagation (corresponding to the entering time and exiting time of the 4-th local mixing state, as well as a longer time). (g)-(i) show the propagation matrices by introducing the new constraint mechanism, corresponding to (d)-(f) respectively.}
\label{example1}
\end{figure}
with the number of layers, we take a simple Newman artificial network \cite{20} as an example. The network consists of 128 nodes divided into four categories of 32 nodes. Each node has on average 14 edges connecting to nodes of the same category and 2 edges connecting to nodes of other categories, as shown in Fig. 6(a). For this four-classification problem, we first calculate the spectrum of its Markov generator (Fig. 6(b)), and then derive the entering time and exiting time of the 4-th local mixing state, i.e., \url{~}2 and \url{~}6, corresponding to the floor and ceiling of the optimal number of layers (Fig. 6(c)). Figs. 6(d), (e) and (f) show the propagation matrices of 2, 6 and 10 steps (or layers) of random walk. As shown, while the propagation matrices between the 2-th and 6-th layers are relatively clear, some impurity information is still introduced. But with the increase of propagation layers, e.g., reaching 10 layers, it will become hard to filter impurity information any more. However, after introducing the constraint mechanism, the propagation matrices of the 2-th and 6-th layers are much clearer (Figs. 6(g) and (h)). More importantly, it will almost not introduce impurity information with the increase of layers, e.g., reaching 10 layers as shown in Fig. 6(i). This further verifies that the new constrained Markov dynamics can suppress the integration of impurity information when propagation, making it more robust and effective.

\subsubsection{Discriminative Aggregation}
After the $k$-step propagation above, we then perform a discriminative aggregation, which forms the relaxation and improvement of the naive model. To be specific, we use the same aggregation as the naive model while aggregating embeddings of the $k$-step propagated neighbors. Then, the final embeddings can be defined in one time as:
\begin{equation}
H^{(k)} = \sigma ((S^{(k)} \circ H^{(0)})W)
\end{equation}
While the model may not distinguish information from different meta-paths in propagation, it does distinguish them in aggregation, achieving the essential selection of different meta-paths. In this way, we can further solve the inherent limitation of the naive model (the difficulty of concatenation in the new aggregation because most nodes do not have the same and complete types of one-hop meta-paths), since we can often get the complete types of neighbors after some $k$ steps of constraint propagation.

Here, one may also concern that the propagation matrix $S^{(k)}$ may become very dense in this case, making the propagation introduce too much noise. But in fact, it is not this case. Thanks to the new constraint mechanism, our $S^{(k)}$ can still remain sparse. Here take a first node $v_1$ in the first category of a complex Lancichinetti artificial network as an example (Fig. 7(a)). After many steps (e.g., $k$ = 10) of propagation, when using the unconstraint random walk, the propagation probability of this node to all the other 999 nodes are positive, showing a dense result (Fig. 7(b)). However, the propagation probability produced by our constraint walk is still sparse (Fig. 7(c)). As shown in Fig. 7(c), our propagation probability of $v_1$ to 766 out of the total 999 are 0; while that to the other nodes are positive. Moreover, the red values (the probability of $v_1$ to nodes in the same category) are often much larger than the blue values (the probability of $v_1$ to nodes outside this category). This demonstrates that our new propagation mechanism can not only obtain a sparse propagation matrix, but also well filter impurity information, making the propagation more effective.

\begin{figure}[htp]
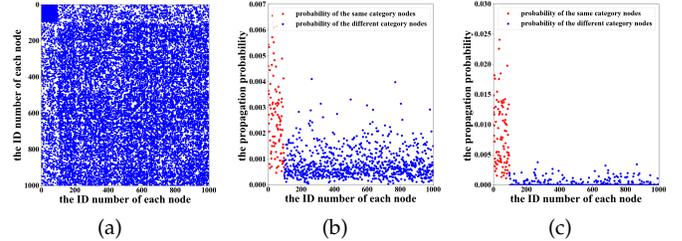

\centering
\subfigbottomskip=0.3pt
\setlength{\abovecaptionskip}{0.13cm}
\subfigure[]{
\includegraphics[width=0.305\linewidth]{point_1.pdf}
}
\subfigure[]{
\label{unconstraint}
\includegraphics[width=0.305\linewidth]{point_2.pdf}
}
\subfigure[]{
\label{constraint}
\includegraphics[width=0.305\linewidth]{point_3.pdf}
}
\caption{\label{network}An Example of illustrating the sparsity of the propagation matrix using our constraint propagation. (a) show an artificial network of 1000 nodes with power-law distribution of degree and category size, generated by Lancichinetti’s model \cite{19}. Here we only use the first category with 97 nodes which are put on the top of the node sequences. We focus on the first node $v_1$ in this category with maximum degree. (b) shows the propagation probability of node $v_1$ to others based on the unconstrained random walk, and (c) that using our constrained walk. Red points denote probabilities of node $v_1$ to nodes in the same category and blue points outside.}
\end{figure}

We define the loss function by using cross entropy as:
\begin{equation}
L=- \sum\limits_{ l\in y_L}Y^lln(C \cdot H^l  )
\end{equation}
where $C$ denotes the set of parameters of the classifier, $y_L$ the set of node indices that have labels, $Y^l$ and $H^l$ the labels and embeddings of the labeled nodes. We use back propagation and Adam optimizer to optimize the model.

\subsection{Implementation}
It is also quite easy to introduce some tricks when implementing our method. The tricks include, for example, supporting the use of candidate meta-path sets and the (multi-head) node-level attention, which are often used in the existing HIN embedding approaches.

First, existing HIN embedding methods often need to use a candidate meta-path set. To make our method support this option, we can adopt only the meta-paths in this candidate set to construct the $k$-step propagation matrix, and then use an aggregation to fuse information from these $k$-step propagated neighbors to derive the final embeddings. 

Second, existing graph neural network-based HIN embedding methods usually adopt the node-level attention for fine-tuning. Our method can also introduce the node-level attention, working together with its inherent algorithmic mechanism of implicitly selecting meta-paths, to further improve performance. To be specific, given a node pair ($u$, $v$) and a specified meta-path $m$, the importance coefficient between nodes $u$ and $v$ can be formulated as:
\begin{equation}
e_{uv}^ m =  \operatorname{LeakyReLU}(\mu_m^T [Wh_u||Wh_v] )
\end{equation}
where $\mu_m$ is the parameterized attention vector for meta-path $m$, and $W$ the mapping matrix applied to each node. After obtaining the importance between nodes $u$ and $v$, we can then use softmax to normalize them to get the weight coefficient as: 
\begin{equation}
 \alpha _{uv}^ m = \operatorname{softmax}_v( e_{uv}^ m)={\operatorname{exp}(e_{uv}^m)\over  \sum_{r \in N_u^m} \operatorname{exp}(e_{ur}^m)}
\end{equation}
Then, the embedding of node $u$ for meta-path $m$ can be aggregated by the neighbor’s embeddings with its corresponding weight coefficients as:
\begin{equation}
h_{u}^m= \sigma ( \sum\limits_{v \in N_u^m} \alpha _{uv}^m Wh_v)
\end{equation}

Finally, we can also extend the node-level attention to a multi-head attention, as done in many existing methods \cite{7}\cite{12}, in order to stabilize the learning process and reduce the high variance (brought by the heterogeneity of networks). That is, we repeat the node-level attention $K$ times, and then concatenate their output as the final embeddings:
\begin{equation}
h_{u}^m=\mathop{\parallel}_{k=1}^K \sigma ( \sum\limits_{v \in N_u^m}\alpha_{uv}^m Wh_v)
\end{equation}

\section{Experiments}
We first give the experimental setup, and then compare our GIAM with some state-of-the-art methods on three network analysis tasks, i.e., node classification, node clustering and network visualization. We finally give an in-depth analysis of different components of our new approach.
\subsection{Experimental Setup}
\subsubsection{Datasets}
We adopt two widely-used heterogeneous information networks from different domains, as shown in Table 2, to evaluate the performance of different methods.
\begin{itemize}[leftmargin=*]
\item \textbf{IMDB} is an online database about TV shows and movie productions. We extract a subset of IMDB with 4278 movies (M), 2081 directors (D) and 5257 actors (A). The movies are divided into three classes (\emph{Action, Comedy, Drama}) based on their genre. Each movie is described by a bag-of-words representation of its plot keywords. The same to \cite{12}, we use the candidate meta-path set \{MAM, MDM\} for algorithms that require such information, and select 400, 400 and 3478 movies as training, validation and testing sets, for semi-supervised learning. 

\renewcommand\arraystretch{1.5}
\begin{table}[!ht]
	\centering
	\caption{\label{table:dataset} Datasets description.}
	\begin{small}
	\resizebox{\linewidth}{!}{
		\begin{tabular}{|c|c|c|c|}
			\hline
				{{Datasets}} &{{No. of Nodes}} & {{No. of Edges}} & {{Meta-paths}} \\ 
			\hline
				\multirow{3}*{IMDB} & 	\multirow{3}*{\shortstack{\#movie(M): 4278\\ \#director(D): 2081 \\
				\#actor (A): 5257}} &
				\multirow{3}*{\shortstack{\#M-D: 4278 \\\#M-A: 12828}} &
				\multirow{3}*{\shortstack{MDM\\MAM}}
				\\
				{} & {} & {} & {} \\
				{} & {} & {} & {} \\
			\hline
				\multirow{4}*{DBLP} & 	\multirow{4}*{\shortstack{\#author (A): 4057\\ \#paper (P): 14328 \\
				\#term (T): 7723\\ \#venue (V): 20}} &
				\multirow{4}*{\shortstack{\#A-P: 19645 \\ \#P-T: 85810 \\ \#P-V: 14328}} &
	       	\multirow{4}*{\shortstack{APA\\APTPA\\APVPA}}
				\\
				{} & {} & {} & {} \\
				{} & {} & {} & {} \\
				{} & {} & {} & {} \\
			\hline
	    \end{tabular}
	}
	\end{small}
\end{table}

\item \textbf{DBLP} is a computer English literature database with authors as its core. We extract a subset of DBLP with 4057 authors (A), 14328 papers (P), 7723 terms (T) and 20 venues (V). The authors are divided into four classes (\emph{Database, Data Mining, Artificial Intelligence and Information Retrieval}) based on their research areas. Each author is described by a bag-of-words representation of his/her paper keywords. Also the same to \cite{12}, we adopt the candidate meta-path set \{APA, APCPA, APTPA\}, and select 400, 400 and 3257 authors as training, validation and testing sets.
\end{itemize}

\subsubsection{Baselines}
We compare our new approach GIAM with eight existing methods. They include: 1) the homogeneous network embedding methods DeepWalk\cite{5}, Node2vec\cite{6}, GCN\cite{2} and GAT\cite{4}, and 2) the HIN embedding methods Metapath2vec\cite{8}, HetGNN\cite{10}, HAN\cite{7} and MAGNN\cite{12}. Especially, GCN is the base of our approach GIAM, and HAN and MAGNN are the state-of-the-art graph neural network-based HIN embedding methods which adopts the hierarchical attention structure. Also of note, we use homogeneous network embedding methods on the HIN structure directly by ignoring the difference of types of nodes and edges.

\renewcommand\arraystretch{1.5}
\begin{table*}[!ht]
	\centering
	\caption{\label{table:classification} Comparisons on node classification.}
	\begin{small}
	\setlength{\tabcolsep}{1.4mm}{
		\begin{tabular}{|c|c|c|c|c|c|c|c|c|c|c|c|}
			\hline
				{{Datasets}} &{{Metrics}} & {{Training ratio}} & {{Deepwalk}} & {{Node2vec}} & {{GCN}} & {{GAT}} & {{Metapath2vec}} & {{HetGNN}} & {{HAN}} & {{MAGNN}} & {{GIAM}} \\ 
			\hline
			\multirow{12}*{IMDB} & \multirow{6}*{\shortstack{Macro-F1\\(\%)}} & 5\% & 41.52 & 43.56 & 54.56 & 54.79 & 42.95 & 42.93 & 55.94 & 54.41 & \textbf{58.49}\\\cline{3-12}
			{} & {} & 10\% & 44.40 & 46.40 & 55.75 & 55.69 & 43.90 & 45.94 & 56.41 & 56.43 & \textbf{59.15}
	        \\ \cline{3-12}
	        {} & {} & 20\% & 46.60 & 49.61 & 56.29 & 56.38 & 45.53 & 48.87 & 57.64 & 57.41 & \textbf{59.79}
	        \\ \cline{3-12}
	        {} & {} & 40\% & 47.92 & 50.87 & 56.00 & 56.26 & 46.39 & 51.39 & 58.46 & 58.70 & \textbf{59.85}
	        \\ \cline{3-12}
	        {} & {} & 60\% & 48.66 & 51.79 & 55.83 & 56.05 & 47.80 & 52.70 & 58.73 & 58.97 & \textbf{60.25}
	        \\ \cline{3-12}
	        {} & {} & 80\% & 48.73 & 52.08 & 56.30 & 56.03 & 48.63 & 53.31 & 58.82 & 59.65 & \textbf{59.97}
	        \\ \cline{2-12}
	        {} & \multirow{6}*{\shortstack{Micro-F1\\(\%)}} & 5\% & 42.31 & 44.13 & 55.22 & 55.48 & 44.31 & 43.80 & 56.28 & 54.61 & \textbf{59.03}
	        \\\cline{3-12}
	        {} & {} & 10\% & 45.45 & 47.32 & 56.23 & 56.20 & 45.75 & 46.89 & 56.62 & 56.59 & \textbf{59.50}
	        \\\cline{3-12}
	        {} & {} & 20\% & 47.88 & 50.59 & 56.58 & 56.60 & 47.06 & 49.62 & 57.66 & 57.43 & \textbf{59.96}
	        \\\cline{3-12}
	        {} & {} & 40\% & 49.47 & 52.01 & 56.39 & 56.52 & 48.12 & 52.24 & 58.46 & 58.85 & \textbf{60.05}
	        \\\cline{3-12}
	        {} & {} & 60\% & 50.20 & 52.92 & 56.19 & 56.31 & 49.50 & 53.58 & 58.75 & 59.09 & \textbf{60.44}
	        \\\cline{3-12}
	        {} & {} & 80\% & 50.33 & 53.45 & 56.52 & 56.14 & 50.65 & 54.40 & 58.95 & 59.76 & \textbf{60.18}
	        
	        \\\hline
			\multirow{12}*{DBLP} & \multirow{6}*{\shortstack{Macro-F1\\(\%)}} & 5\% & 73.09 & 78.02 & 85.59 & 79.67 & 90.17 & 90.83 & 91.80 & 92.96 & \textbf{93.24}
			\\\cline{3-12}
			{} & {} & 10\% & 80.95 & 84.53 & 86.11 & 84.99 & 90.76 & 91.18 & 92.27 & 93.07 & \textbf{93.48}
	        \\ \cline{3-12}
	        {} & {} & 20\% & 84.08 & 85.51 & 86.88 & 86.72 & 91.28 & 91.68 & 92.88 & 92.92 & \textbf{93.64}
	        \\ \cline{3-12}
	        {} & {} & 40\% & 86.98 & 86.82 & 88.12 & 87.57 & 91.88 & 92.20 & 93.03 & 93.17 & \textbf{93.76}
	        \\ \cline{3-12}
	        {} & {} & 60\% & 88.59 & 88.14 & 87.84 & 88.32 & 92.31 & 92.36 & 92.97 & 93.50 & \textbf{93.70}
	        \\ \cline{3-12}
	        {} & {} & 80\% & 89.99 & 88.78 & 87.75 & 89.16 & 92.70 & 92.22 & 93.18 & 93.52 & \textbf{93.96}
	        \\ \cline{2-12}
	        {} & \multirow{6}*{\shortstack{Micro-F1\\(\%)}} & 5\% & 75.49 & 80.41 & 86.08 & 82.88 & 90.90 & 91.39 & 92.36 & 93.49 & \textbf{93.72}
	        \\\cline{3-12}
	        {} & {} & 10\% & 81.96 & 85.46 & 86.62 & 86.02 & 91.43 & 91.74 & 92.81 & 93.58 & \textbf{93.96}
	        \\\cline{3-12}
	        {} & {} & 20\% & 85.02 & 86.48 & 87.28 & 87.38 & 91.97 & 92.20 & 93.36 & 93.43 & \textbf{94.12}
	        \\\cline{3-12}
	        {} & {} & 40\% & 87.81 & 87.68 & 88.50 & 88.18 & 92.50 & 92.68 & 93.50 & 93.63 & \textbf{94.23}
	        \\\cline{3-12}
	        {} & {} & 60\% & 89.38 & 89.02 & 88.28 & 88.98 & 92.90 & 92.88 & 93.47 & 93.95 & \textbf{94.18}
	        \\\cline{3-12}
	        {} & {} & 80\% & 90.43 & 89.51 & 88.16 & 89.69 & 93.25 & 92.78 & 93.67 & 93.96 & \textbf{94.39}
	        \\\hline
		\end{tabular}
	}
	\end{small}
 	\vspace{-0.1cm}
\end{table*}

\subsubsection{Parameter Settings}
For the methods based on semi-supervised graph neural networks (including GCN, GAT, HAN, MAGNN and our GIAM), we set the dropout rate to 0.5 and use the same splits for training, verification and testing sets. We employ the Adam optimizer with the learning rate setting to 0.005 and apply early stopping with a patience of 50. For GAT, HAN and MAGNN, we set the number of attention heads to 8. For HAN and MAGNN, we set the dimension of the meta-path-level attention vector to 128. For the methods based on random walk (including DeepWalk, Node2vec, HetGNN and metapath2vec), we set the window size to 5, walk length to 100, walks per node to 40, and the number of negative samples to 5. For a fair comparison, the embedding dimension of all methods mentioned above is set to 64.

\subsection{Comparisons to Existing Methods}
We first make a quantitative comparison on node classification and clustering, and then a qualitative comparison on visualization.

\subsubsection{Node Classification}
On the node classification task, for each method, we first generate the embeddings of the labeled nodes
(i.e., movies in IMDB and authors in DBLP), and then feed them to SVM by using different training ratios from 5\% to 80\% (as done in the most existing works). Since the variance of the graph structure data can be quite large, we repeat this process 10 times and report the average \emph{Macro-F1} and \emph{Micro-F1}.

The results are shown in Table 3. As shown, the proposed method GIAM always performs the best across different training ratios and datasets. On the IMDB dataset, GIAM is 1.15-2.88\% and 0.32-4.42\% more accurate than the best baselines HAN and MAGNN, which are also the heterogeneous graph neural network methods (while they use mate-path-level attentions directly). On the DBLP dataset, GIAM is 0.71-1.44\% and 0.20-0.72\% more accurate than the best baselines HAN and MAGNN in the case of an already very high base accuracy ($\geq$ 91.80\%), making our improvement still nontrivial. These results not only demonstrate the superiority of the new propagation and aggregation mechanism, but also validate the effectiveness of our main idea of using algorithmic mechanisms (rather than the meta-path-level attention directly) to implicitly achieve the role of attention of selecting meta-paths. In addition, the performance of GIAM is much better than that of GCN (i.e., 3.27-4.42\% and 5.64-7.65\% more accurate than IMDB and DBLP), which further demonstrates the effectiveness of our new mechanism for distinguishing importance of information with respect to different meta-paths in both propagation and aggregation. 

\subsubsection{Node Clustering}
We also conduct comparisons of these methods on node clustering.
\renewcommand\arraystretch{1.5}
\begin{table*}[htp]
	\centering
	\caption{\label{table:NMI} Comparisons on node clustering in terms of NMI. AVG shows the average result.}
	\begin{small}
	\setlength{\tabcolsep}{1.97mm}{
		\begin{tabular}{|c|c|c|c|c|c|c|c|c|c|}
			\hline
			\multirow{2}*{Datasets}&\multicolumn{9}{|c|}{NMI (\%)}
			\\ \cline{2-10}
			&Deepwalk & Node2vec & GCN & GAT & Metapath2vec    	&HetGNN & HAN &MAGNN	&GIAM\\
			\hline
			IMDB & 0.55 & 5.34 & 10.42 & 10.02 & 0.43 & 0.46 & 13.02 & 13.77 & \textbf{15.41}\\
			\hline
			DBLP & 71.78 & 74.80 & 53.93 & 68.15 & 75.02 & 74.26 & 73.13 & \textbf{78.97} & 78.27 (2)\\
			\hline
			AVG & 36.17 & 40.07 & 32.18 & 39.09 & 37.73  & 37.36 & 43.08 & 46.37 & \textbf{46.84}\\
			\hline
	\end{tabular}
	}
	\end{small}
\end{table*}
\renewcommand\arraystretch{1.5}
\begin{table*}[htp]
	\centering
	\caption{\label{table:ARI}Comparisons on node clustering in terms of ARI.}
	\begin{small}
	\setlength{\tabcolsep}{1.7mm}{
		\begin{tabular}{|c|c|c|c|c|c|c|c|c|c|}
			\hline
			\multirow{2}*{Datasets}&\multicolumn{9}{|c|}{ARI [-1,1]}
			\\ \cline{2-10}
			&Deepwalk & Node2vec & GCN & GAT & Metapath2vec  & HetGNN & HAN & MAGNN & GIAM\\
			\hline
			IMDB & -0.0014 & 0.0642 & 0.0661 & 0.0744 & 0.0005 & 0.0048 & 0.1282 & 0.1206 & \textbf{0.1552}\\
			\hline
			DBLP & 0.7415 & 0.7796 & 0.4670 & 0.6859 & 0.7945 & 0.8028 & 0.7938 & \textbf{0.8392} & 0.8273 (2)\\
			\hline
			AVG & 0.3701 & 0.4219 & 0.2666 & 0.3802 & 0.3975 & 0.4038 & 0.4610 & 0.4799 & \textbf{0.4913}\\
			\hline
	\end{tabular}
	}
	\end{small}
 	\vspace{-0.1cm}
\end{table*}
In this task, for each method, we first generate embeddings of the labeled nodes, and then feed them to K-Means algorithm. The number of clusters $K$ is set to the same as the ground-truth, i.e., 3 for IMDB and 4 for DBLP. Since the performance of K-Means is easily affected by the initial center, we repeat the process 10 times and report the average \emph{normalized mutual information} (NMI) and \emph{adjusted rand index} (ARI).

The results are shown in Tables 4 and 5. As shown, the proposed method GIAM performs the best on IMDB. While GIAM performs the second best on DBLP, its performance is still very competitive with that of the best baseline MAGNN. On average on both these two datasets, GIAM is 10.67\%, 6.77\%, 14.66\%, 7.75\%, 9.11\%, 9.48\%, 3.76\% and 0.47\% more accurate than Deepwalk, Node2vec, GCN, GAT, Metapath2vec, HetGNN, HAN and MAGNN in terms of NMI; and 0.1212, 0.0694, 0.2247, 0.1111, 0.0938, 0.0875, 0.0303 and 0.0114 better than these methods in ARI (in the range of -1 to 1). Moreover, (on average) GIAM is still better than the methods using meta-path-level attentions directly (i.e., HAN and MAGNN). This further validates the soundness of using algorithmic mechanisms to evaluate importance of different meta-paths. Neither GCN nor GAT is so competitive here. This is mainly because they fail to distinguish importance of information with respect to different meta-paths, which significantly compromises their performance in the unsupervised clustering setting.

\subsubsection{Visualization}
For a more intuitively comparison, we also visualize the embeddings of author nodes of some representative network embedding methods (i.e., GCN, HetGNN, HAN and our GIAM) on the DBLP dataset as an example. We utilize the well-known t-SNE tool \cite{29} to project node 
\begin{figure}[htp]
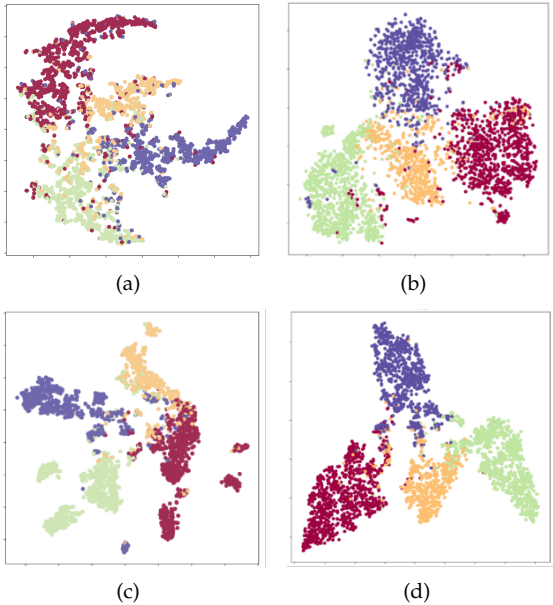

\setlength{\abovecaptionskip}{0.13cm}
\subfigbottomskip=0.5pt
\centering
\subfigure[]{
\label{GCN}
\includegraphics[width=0.39\linewidth]{GCN.pdf}
}
\subfigure[]{
\includegraphics[width=0.4\linewidth]{HetGNN.pdf}
}
\\
\subfigure[]{
\includegraphics[width=0.4\linewidth]{HAN.pdf}
}
\subfigure[]{
\includegraphics[width=0.4\linewidth]{GIAM.pdf}
}
\caption{\label{tsne}The visualization of author nodes of the embeddings learned by (a) GCN, (b) HetGNN, (c) HAN and (d) GIAM on DBLP. Different colors correspond to different research areas in ground truth.}
\end{figure}
embeddings to two dimensions. Different colors correspond to different research areas of these nodes.

As shown in Fig. \ref{tsne}, GCN (which ignores the heterogeneity of nodes) does not perform well, i.e., the author nodes belong to different research areas are sometimes mixed with each other. HetGNN performs much better than GCN, but its boundary is still blurry. While both HAN and our GIAM separate the author nodes in different research areas reasonably well, our GIAM has a more distinct boundary and denser cluster structures in visualization.

\subsection{A Deep Analysis of GIAM}
Similar to most deep learning models, GIAM also contains some important components that may have significant impact on the performance. To test the effectiveness of each component of GIAM, we conduct experiments on comparing GIAM with four variations. The variants are as follows: 1) GCN which serves as the base framework of GIAM of not distinguishing importance of information with respect to different meta-paths, 2) the naive model of GIAM, named as GIAM-1, 3) GIAM of removing node-level attention (by assigning the same importance to each neighbor node), named as GIAM-2, and 4) GIAM of adding the meta-path-level attention, named as GIAM-3. We take their comparison on node classification as an example.

As shown in Table 6, compared to GCN, the naive model GIAM-1 (which distinguishes meta-paths) has an obvious improvement, i.e., 0.86-1.15\% and 4.18-5.25\% more accurate on IMDB and DBLP. However, due to the sparsity of HINs, GIAM-1 inevitably needs to add a large number of non-informative features, so as to fill in embeddings of the missing types of one-hop meta-paths during aggregation. While its result is basically satisfactory, this limitation compromises performance inevitably. We overcome this limitation by introducing a new mechanism of relaxation and improvement, deriving GIAM-2, which further improves performance of the naive model, i.e., 2.35-3.63\% and 0.02-2.76\% more accurate on IMDB and DBLP. Furthermore, by introducing the fine-turning node-level attention, the derived GIAM improves GIAM-2 on DBLP (i.e., 0.63-0.87\% more accurate), while the improvement on IMDB is not so obvious (because IMDB is harder to be trained well with a relative low accuracy, easier leading to overfitting). This further demonstrates that the node-level attention indeed plays a fine-tuning role when the model can be well trained (such as on DBLP with a relative high accuracy). Finally, GIAM-3 of adding the meta-path-level attention hardly changes the performance of GIAM. This further validates that our algorithmic mechanism has already played a significant role in selecting meta-paths, compared to the explicit meta-path-level attention approach.
\renewcommand\arraystretch{1.5}
\begin{table}[h]
    \caption{Comparisons of our GIAM with four variants
(GCN, GIAM-1, GIAM-2 and GIAM-3) on node classification.}
	\label{component}
	\begin{small}
	
  	\resizebox{\linewidth}{!}{
		\begin{tabular}{|c|c|c|c|c|c|c|c|}
			\hline
				{{Datasets}} &{{Metrics}} & {{Training radio}} & {{GCN}}& {{GIAM-1}} & {{GIAM-2}} & {{GIAM}} & {{GIAM-3}}  \\ 
			\hline
			\multirow{12}*{IMDB} & \multirow{6}*{\shortstack{Macro-F1\\(\%)}} & 5\% & 54.56 & 55.52 & 58.29 & 58.49 & 58.56 
			\\ \cline{3-8}
			{} & {} & 10\% & 55.75 & 56.73 & 59.31 & 59.15 & 59.26
	        \\ \cline{3-8}
	        {} & {} & 20\% & 56.29 & 57.15 & 59.90 & 59.79 & 59.94
	        \\ \cline{3-8}
	        {} & {} & 40\% & 56.00 & 56.99 & 60.01 & 59.93 & 59.85
	        \\ \cline{3-8}
	        {} & {} & 60\% & 55.83 & 56.88 & 60.51 & 60.25 & 60.31
	        \\ \cline{3-8}
	        {} & {} & 80\% & 56.30 & 57.34 & 60.43 & 59.97 & 60.10 
	        \\ \cline{2-8}
	        {} & \multirow{6}*{\shortstack{Micro-F1\\(\%)}} & 5\% & 55.22& 56.14 & 58.80 & 59.03 & 59.11
	        \\\cline{3-8}
	        {} & {} & 10\% & 56.23 & 57.20 & 59.55 & 59.50 & 59.61	        
	        \\\cline{3-8}
	        {} & {} & 20\% & 56.58 & 57.51 & 59.96 & 59.96 & 60.10
	        \\\cline{3-8}
	        {} & {} & 40\% & 56.39 & 57.48 & 60.10 & 60.05 & 60.13
	        \\\cline{3-8}
	        {} & {} & 60\% & 56.19 & 57.34 & 60.55 & 60.44 & 60.50 
	        \\\cline{3-8}
	        {} & {} & 80\% & 56.52 & 57.66 & 60.47 & 60.18 & 60.30 
	        
	        \\\hline
			\multirow{12}*{DBLP} & \multirow{6}*{\shortstack{Macro-F1\\(\%)}} & 5\% & 85.59 & 89.77 & 92.53 & 93.24 & 93.25 
			\\\cline{3-8}
			{} & {} & 10\% & 86.11 & 90.85 & 92.62  & 93.48 & 93.48
	        \\ \cline{3-8}
	        {} & {} & 20\% & 86.88 & 91.89 & 92.79 & 93.64 & 93.61 	  
	        \\ \cline{3-8}
	        {} & {} & 40\% & 88.12 & 92.41 & 92.89 & 93.76  & 93.78 
	        \\ \cline{3-8}
	        {} & {} & 60\% & 87.84 & 92.81 & 92.87  & 93.70 & 93.69
	        \\ \cline{3-8}
	        {} & {} & 80\% & 87.75 & 91.98 & 93.12  & 93.96 & 93.98
	        \\ \cline{2-8}
	        {} & \multirow{6}*{\shortstack{Micro-F1\\(\%)}} & 5\% & 86.08 & 90.58 & 93.09 & 93.72 & 93.75
	        \\\cline{3-8}
	        {} & {} & 10\% & 86.62 & 91.57 & 93.17 & 93.96 & 93.96 
	        \\\cline{3-8}
	        {} & {} & 20\% & 87.28 & 92.53 & 93.35  & 94.12 & 94.10
	        \\\cline{3-8}
	        {} & {} & 40\% & 88.50 & 93.01 & 93.45 & 94.23 & 94.26
	        \\\cline{3-8}
	        {} & {} & 60\% & 88.28 & 93.42 & 93.44 & 94.18 & 94.19
	        \\\cline{3-8}
	        {} & {} & 80\% & 88.16 & 92.52 & 93.66 & 94.39 & 94.43
	        \\\hline
		\end{tabular}
  	}
	\end{small}
 	\vspace{-0.4cm}
\end{table}

\section{Related Work}
Heterogeneous information network (HIN) embedding aims to learn a low-dimensional distributed representation for each node of a HIN while preserving the structure and semantic information. Existing HIN embedding methods can be mainly divided into three categories, including the random walk-based methods, the relation learning based methods and the graph neural network-based methods.

The random walk-based methods first utilize random walk on a HIN to generate the node walk sequences, and then feed them to the subsequent model to obtain node embeddings. For example, JUST\cite{3} adopts the jump and stay strategies on a HIN, which select the next node based on the probability of the jump or stay operation, to perform random walk. It then inputs the generated walk sequences to the skip-gram model to obtain the final node embeddings. Metapath2vec\cite{8} first generates the node walk sequences based on meta-paths, and then obtains the node embeddings by adopting heterogeneous skip-gram with negative sampling. HetGNN \cite{10} improves metapath2vec by incorporating additional node information. It first introduces a sampling strategy based on random walk with restart to sample neighbors for each node, and then uses a heterogeneous neural network architecture to aggregate the feature information of those sampled neighbor nodes.

The relation learning based methods aim to learn a scoring function which evaluates an arbitrary triplet composed of two nodes and an edge type, and output a scalar to measure the acceptability of this triplet. For example, DistMult\cite{15} adopts a similarity-based scoring function to learn the edge possibility between arbitrary two nodes of the HIN. ConvE\cite{16} proposes a deep neural model instead of the simple similarity function to score the edge possibility between two nodes. TransE\cite{28} learns the edge possibility between two nodes by using a translational distance.

The graph neural network-based methods aim to learn node embeddings by aggregating the information from neighbor nodes of a HIN. For example, HAN\cite{7} proposes a hierarchical attention mechanism, including the node-level and semantic-level attentions, to aggregate the information from meta-path-based neighbors. To be specific, node-level attention learns the importance of neighbors in the same meta-path while semantic-level attention learns the importance of different meta-paths. MAGNN\cite{12} employs three major components, i.e., the node-type specific transformation, the node-level meta-path instance aggregation and the meta-path-level embedding fusion, to obtain the node embeddings of heterogeneous graphs. While those graph neural network-based methods can often derive satisfactory node embeddings, they still have some essential limitations. That is, the complicated hierarchical attention structure often makes these methods difficult to really achieve the goal of selecting meta-paths, partly due to the highly overfitting (as shown in Fig. 1(a) as an illustrative example). Meanwhile, those methods treat the one-hop and multi-hop meta-paths indistinguishably to propagate information, which may be not so intuitive from the perspective of network propagation dynamics in network science. 

\section{Conclusion}
We propose a novel GCN-based method, namely GIAM, via implicitly (rather than explicitly) utilizing attention and meta-paths, in order to effectively achieve HIN embedding. We use the direct linked meta-paths, a discriminative aggregation, along with the stacked layers of propagation, to distinguish the importance of different meta-paths. We further give an effective relaxation and improvement by introducing a new multi-layer propagation which is separated from the aggregation. That is, we first replace the spectral filter of GCN from the symmetric normalized graph Laplacian to an equivalent asymmetric one and remove activation functions, making it a well-defined probabilistic propagation process. We then introduce a random graph-based constraint mechanism RPC on this probabilistic propagation, to avoid importing too much noise with the increase of propagation layers. Empirical results on various graph mining tasks, including node classification, node clustering and graph visualization, demonstrate the superiority of our new approach over some state-of-the-art methods.

\ifCLASSOPTIONcaptionsoff
  \newpage
\fi

\bibliographystyle{ieeetr} 
\bibliography{ijcai20}

\begin{thebibliography}{10}

\bibitem{17}
C.~Yang, Y.~Xiao, Y.~Zhang, Y.~Sun, and J.~Han, ``Heterogeneous network
  representation learning: Survey, benchmark, evaluation, and beyond,'' {\em
  CoRR}, vol.~abs/2004.00216, 2020.

\bibitem{23}
C.~Shi, Y.~Li, J.~Zhang, Y.~Sun, and P.~S. Yu, ``A survey of heterogeneous
  information network analysis,'' {\em IEEE Transactions on Knowledge and Data
  Engineering}, vol.~29, no.~1, pp.~17--37, 2017.

\bibitem{27}
W.~Shen, J.~Han, J.~Wang, X.~Yuan, and Z.~Yang, ``{SHINE+:} {A} general
  framework for domain-specific entity linking with heterogeneous information
  networks,'' {\em {IEEE} Transactions on Knowledge and Data Engineering.},
  vol.~30, no.~2, pp.~353--366, 2018.

\bibitem{25}
B.~Hu, C.~Shi, W.~X. Zhao, and P.~S. Yu, ``Leveraging meta-path based context
  for top-{N} recommendation with {A} neural co-attention model,'' in {\em
  Proceedings of SIGKDD, ACM, 2018}, pp.~1531--1540.

\bibitem{26}
C.~Wang, Y.~Song, H.~Li, M.~Zhang, and J.~Han, ``Unsupervised meta-path
  selection for text similarity measure based on heterogeneous information
  networks,'' {\em Data Mining and Knowledge Discovery}, vol.~32, no.~6,
  pp.~1735--1767, 2018.

\bibitem{21}
C.~Park, D.~Kim, J.~Han, and H.~Yu, ``Unsupervised attributed multiplex network
  embedding,'' in {\em Proceedings of AAAI, 2020}, pp.~5371--5378.

\bibitem{24}
W.~L. Hamilton, Z.~Ying, and J.~Leskovec, ``Inductive representation learning
  on large graphs,'' in {\em Proceedings of NIPS, 2017}, pp.~1024--1034.

\bibitem{3}
R.~Hussein, D.~Yang, and P.~Cudr{\'{e}}{-}Mauroux, ``Are meta-paths necessary?:
  Revisiting heterogeneous graph embeddings,'' in {\em Proceedings of CIKM,
  2018}, pp.~437--446.

\bibitem{8}
Y.~Dong, N.~V. Chawla, and A.~Swami, ``metapath2vec: Scalable representation
  learning for heterogeneous networks,'' in {\em Proceedings of SIGKDD, ACM,
  2017}, pp.~135--144.

\bibitem{15}
B.~Yang, W.~Yih, X.~He, J.~Gao, and L.~Deng, ``Embedding entities and relations
  for learning and inference in knowledge bases,'' in {\em Proceedings of ICLR,
  2015}.

\bibitem{16}
T.~Dettmers, P.~Minervini, P.~Stenetorp, and S.~Riedel, ``Convolutional 2d
  knowledge graph embeddings,'' in {\em Proceedings of AAAI, 2018},
  pp.~1811--1818.

\bibitem{7}
X.~Wang, H.~Ji, C.~Shi, B.~Wang, Y.~Ye, P.~Cui, and P.~S. Yu, ``Heterogeneous
  graph attention network,'' in {\em Proceedings of WWW, 2019}, pp.~2022--2032.

\bibitem{12}
X.~Fu, J.~Zhang, Z.~Meng, and I.~King, ``{MAGNN:} metapath aggregated graph
  neural network for heterogeneous graph embedding,'' in {\em Proceedings of
  WWW, 2020}, pp.~2331--2341.

\bibitem{2}
T.~N. Kipf and M.~Welling, ``Semi-supervised classification with graph
  convolutional networks,'' in {\em Proceedings of ICLR, 2017}.

\bibitem{14}
J.~Bruna, W.~Zaremba, A.~Szlam, and Y.~LeCun, ``Spectral networks and locally
  connected networks on graphs,'' in {\em Proceedings of ICLR, 2014}.

\bibitem{1}
M.~Defferrard, X.~Bresson, and P.~Vandergheynst, ``Convolutional neural
  networks on graphs with fast localized spectral filtering,'' in {\em
  Proceedings of NIPS, 2016}, pp.~3837--3845.

\bibitem{32}
Y.~Wang, Z.~Duan, B.~Liao, F.~Wu, and Y.~Zhuang, ``Heterogeneous attributed
  network embedding with graph convolutional networks,'' in {\em Proceedings of
  AAAI, 2019}, pp.~10061--10062.

\bibitem{11}
F.~Wu, A.~H.~S. Jr., T.~Zhang, C.~Fifty, T.~Yu, and K.~Q. Weinberger,
  ``Simplifying graph convolutional networks,'' in {\em Proceedings of ICML,
  2019}, pp.~6861--6871.

\bibitem{30}
B.~Yang, J.~Liu, and J.~Feng, ``On the spectral characterization and scalable
  mining of network communities,'' {\em {IEEE} Transactions on Knowledge and
  Data Engineerin}, vol.~24, no.~2, pp.~326--337, 2012.

\bibitem{31}
M.~E.~J. Newman and M.~Girvan, ``Finding and evaluating community structure in
  networks,'' {\em Physical Review E}, vol.~69, no.~2, pp.~026113--026113,
  2004.

\bibitem{20}
N.~M. E.~J. Girvan~M, ``Community structure in social and biological
  networks,'' {\em Proceedings of the National Academy of Sciences}, vol.~99,
  no.~12, pp.~7821--7826, 2002.

\bibitem{19}
A.~Lancichinetti and S.~Fortunato, ``Benchmarks for testing community detection
  algorithms on directed and weighted graphs with overlapping communities,''
  {\em Physical Review E}, vol.~80, no.~1, p.~016118, 2009.

\bibitem{5}
B.~Perozzi, R.~Al{-}Rfou, and S.~Skiena, ``Deepwalk: online learning of social
  representations,'' in {\em Proceedings of SIGKDD, ACM, 2014}, pp.~701--710.

\bibitem{6}
A.~Grover and J.~Leskovec, ``node2vec: Scalable feature learning for
  networks,'' in {\em Proceedings of SIGKDD, ACM, 2016}, pp.~855--864.

\bibitem{4}
P.~Velickovic, G.~Cucurull, A.~Casanova, A.~Romero, P.~Li{\`{o}}, and
  Y.~Bengio, ``Graph attention networks,'' in {\em Proceedings of ICLR, 2018}.

\bibitem{10}
C.~Zhang, D.~Song, C.~Huang, A.~Swami, and N.~V. Chawla, ``Heterogeneous graph
  neural network,'' in {\em Proceedings of SIGKDD, ACM, 2019}, pp.~793--803.

\bibitem{29}
V.~D.~M. Laurens and G.~Hinton, ``Visualizing data using t-sne,'' {\em Journal
  of Machine Learning Research}, vol.~9, no.~2605, pp.~2579--2605, 2008.

\bibitem{28}
A.~Bordes, N.~Usunier, A.~Garc{\'{\i}}a{-}Dur{\'{a}}n, J.~Weston, and
  O.~Yakhnenko, ``Translating embeddings for modeling multi-relational data,''
  in {\em Proceedings of NIPS, 2013}, pp.~2787--2795.

\end{thebibliography}

\appendix
\section{}
In Section 2, we have used three graph neural network-based HIN embedding methods, i.e., HAN, MAGNN and our new approach GIAM, to conduct the motivating experiment on two widely-used heterogeneous information networks, i.e., IMDB and DBLP. Here, on each network, we give the detailed results of different methods on different radios (i.e., 5-80\%) of the supervised information, as shown in Table 7 and Table 8, respectively.

\renewcommand\arraystretch{1.5}
\begin{table}[h]
    \caption{The performance of HAN and MAGNN of using (and not using) meta-path-level attention, as well as our new approach GIAM on the IMDB dataset. HAN-1 denotes HAN of using meta-path-level attention and HAN-2 not. MAGNN-1 denotes MAGNN of using meta-path-level attention and MAGNN-2 not. AVG shows the average result.}
	\label{component}
	\begin{small}
	
  	\resizebox{\linewidth}{!}{
		\begin{tabular}{|c|c|c|c|c|c|c|c|}
			\hline
				{{Dataset}} &{{Metrics}} & {{Training radio}} & {{HAN-1}}& {{HAN-2}} & {{MAGNN-1}} & {{MAGNN-2}} & {{GIAM}}  \\ 
			\hline
			\multirow{14}*{IMDB} & \multirow{7}*{\shortstack{Macro-F1\\(\%)}} & 5\% & 55.94 & 57.57 & 54.41 & 55.27 & 58.49 
			\\ \cline{3-8}
			{} & {} & 10\% & 56.41 & 58.35 & 56.43 & 56.44 & 59.15
	        \\ \cline{3-8}
	        {} & {} & 20\% & 57.64 & 59.16 & 57.41 & 58.72 & 59.79
	        \\ \cline{3-8}
	        {} & {} & 40\% & 58.46 & 59.49 & 58.70 & 59.71 & 59.85
	        \\ \cline{3-8}
	        {} & {} & 60\% & 58.73 & 59.55 & 58.97 & 59.71 & 60.25
	        \\ \cline{3-8}
	        {} & {} & 80\% & 58.82 & 59.43 & 59.65 & 59.95 & 59.97 
	        \\ \cline{3-8}
	        {} & {} & AVG & 57.67 & 58.93 & 57.60 & 58.30 & 59.58 
	        \\ \cline{2-8}
	        {} & \multirow{7}*{\shortstack{Micro-F1\\(\%)}} & 5\% & 56.28& 57.94 & 54.61 & 55.39 & 59.03
	        \\\cline{3-8}
	        {} & {} & 10\% & 56.62 & 58.53 & 56.59 & 56.71 & 59.50	    
	        \\\cline{3-8}
	        {} & {} & 20\% & 57.66 & 59.21 & 57.43 & 58.83 & 59.96
	        \\\cline{3-8}
	        {} & {} & 40\% & 58.46 & 59.53 & 58.85 & 59.89 & 60.05
	        \\\cline{3-8}
	        {} & {} & 60\% & 58.75 & 59.53 & 59.09 & 59.91 & 60.44 
	        \\\cline{3-8}
	        {} & {} & 80\% & 58.95 & 59.40 & 59.76 & 60.24 & 60.18
	         \\\cline{3-8}
	        {} & {} & AVG & 57.79 & 59.02 & 57.72 & 58.50 & 59.86
	        \\\hline
		\end{tabular}
  	}
	\end{small}
 	\vspace{-0.4cm}
\end{table}

\renewcommand\arraystretch{1.5}
\begin{table}[h]
    \caption{The performance of HAN and MAGNN of using (and not using) meta-path-level attention, as well as our new approach GIAM on the DBLP dataset.}
	\label{component}
	\begin{small}
	
  	\resizebox{\linewidth}{!}{
		\begin{tabular}{|c|c|c|c|c|c|c|c|}
			\hline
				{{Dataset}} &{{Metrics}} & {{Training radio}} & {{HAN-1}}& {{HAN-2}} & {{MAGNN-1}} & {{MAGNN-2}} & {{GIAM}}  \\ 
			\hline
			\multirow{14}*{DBLP} & \multirow{7}*{\shortstack{Macro-F1\\(\%)}} & 5\% & 91.80 & 92.10 & 92.96 & 88.05 & 93.24
			\\ \cline{3-8}
			{} & {} & 10\% & 92.27 & 92.31 & 93.07 & 88.65 &93.48
	        \\ \cline{3-8}
	        {} & {} & 20\% & 92.88 & 92.54 & 92.92 & 89.87 & 93.64
	        \\ \cline{3-8}
	        {} & {} & 40\% & 93.03 & 92.49 & 93.17 & 91.38 & 93.76
	        \\ \cline{3-8}
	        {} & {} & 60\% & 92.97 & 92.66 & 93.50 & 92.29 & 93.70
	        \\ \cline{3-8}
	        {} & {} & 80\% & 93.18 & 92.70 & 93.52 & 92.30 & 93.96
	        \\ \cline{3-8}
	        {} & {} & AVG & 
	        92.69 & 92.47 & 93.19 & 90.42  & 93.63
	        \\ \cline{2-8}
	        {} & \multirow{7}*{\shortstack{Micro-F1\\(\%)}} & 5\% & 
	        92.36 & 92.68 & 93.49 & 88.90 & 93.72
	        \\\cline{3-8}
	        {} & {} & 10\% & 92.81 & 92.89 &  93.58 & 89.44 & 93.96
	        \\\cline{3-8}
	        {} & {} & 20\% & 93.36 & 93.10  & 93.43 & 90.56 & 94.12
	        \\\cline{3-8}
	        {} & {} & 40\% & 93.50 & 93.06 & 93.63 & 91.95 & 94.23
	        \\\cline{3-8}
	        {} & {} & 60\% & 93.47 & 93.24 & 93.95 & 92.81 & 94.18
	        \\\cline{3-8}
	        {} & {} & 80\% & 93.67 & 93.27 & 93.96 & 92.81 & 94.39
	         \\\cline{3-8}
	        {} & {} & AVG & 93.20 & 93.04 & 93.67 & 91.08 & 94.10
	        \\\hline
		\end{tabular}
  	}
	\end{small}
 	\vspace{-0.4cm}
\end{table}

\end{document}